\def\year{2019}\relax
\documentclass[letterpaper]{article} 
\usepackage[table]{xcolor}
\usepackage{aaai19}  
\usepackage{times}  
\usepackage{helvet}  
\usepackage{courier}  
\usepackage{url}  
\usepackage{graphicx}  
\usepackage{bm}
\usepackage{bbm}
\usepackage{subcaption}
\usepackage{caption} 
\usepackage{amsmath,amsfonts,amssymb}
\usepackage{amsthm}
\usepackage{booktabs}
\usepackage{algorithm}
\usepackage[noend]{algpseudocode}
\DeclareMathOperator*{\argmin}{argmin}
\DeclareMathOperator*{\argmax}{argmax}
\newcommand{\norm}[1]{\left\lVert#1\right\rVert}

\begin{document}
%
\title{Noise Contrastive Estimation for Scalable Linear Models\\ for One-Class Collaborative Filtering}
\author{
Ga Wu\textsuperscript{1,2,3}, ~
Maksims Volkovs\textsuperscript{2}, ~
Chee Loong Soon\textsuperscript{1}, ~
Scott Sanner\textsuperscript{1,3}, ~
Himanshu Rai\textsuperscript{2}\\
\textsuperscript{1}{University of Toronto}, ~
\textsuperscript{2}{Layer6 AI}, ~
\textsuperscript{3}{Vector Institute}\\
\{wuga, ssanner\}@mie.utoronto.com,
\{maks, himanshu\}@layer6.ai,
cheeloong.soon@mail.utoronto.ca}
\maketitle
\begin{abstract}
Previous highly scalable one-class collaborative filtering methods such as Projected Linear Recommendation (PLRec) have advocated using fast randomized SVD to embed items into a latent space, followed by linear regression methods to learn personalized recommendation models per user.  Unfortunately, naive SVD embedding methods often exhibit a popularity bias that skews the ability to accurately embed niche items.  To address this, we leverage insights from Noise Contrastive Estimation (NCE) to derive a closed-form, efficiently computable ``depopularized'' embedding.  While this method is not ideal for direct recommendation using methods like PureSVD since popularity still plays an important role in recommendation, we find that embedding followed by linear regression to learn personalized user models in a novel method we call NCE-PLRec leverages the improved item embedding of NCE while correcting for its popularity unbiasing in final recommendations.  An analysis of the recommendation popularity distribution demonstrates that NCE-PLRec uniformly distributes its recommendations over the popularity spectrum while other methods exhibit distinct biases towards specific popularity subranges, thus artificially restricting their recommendations.  Empirically, NCE-PLRec outperforms state-of-the-art methods as well as various ablations of itself on a variety of large-scale recommendation datasets.  
\end{abstract}


\section{Introduction}

In an era of virtually unlimited choices, recommender systems are necessary to assist users in finding items they may like.  Collaborative filtering (CF) is the de-facto standard approach for making such personalized recommendations based on automated collection of item interaction data from a population of users~\cite{sarwar2002recommender}.  However, in many cases, these interactions lack explicit negative signals, e.g., clicks on a website or purchases of a book.  In these cases, a lack of interaction should not be construed as implicitly negative; indeed, it could simply be that a user was unaware of the item's existence.  This recommendation setting where only positive (and typically very sparse) interactions are observed is known as the One Class Collaborative Filtering (OC-CF) problem~\cite{pan2008one}.



One approach to tackle OC-CF is to factorize a large sparse implicit matrix into a smaller latent matrix of user and item representations \cite{pan2008one,hu2008collaborative}. 
However, matrix factorization requires optimizing a non-convex objective, resulting in local optima and the need for substantial hyperparameter tuning for good practical performance~\cite{langville2006initializations}. 
%
An alternative scalable solution is to first reduce the dimensionality of the matrix, then learn the importance of different latent projected features using linear regression. Methods such as~\cite{sanner:ijcai16a}, which we refer to as Projected Linear Recommendation (PLRec) precompute the item embeddings through fast randomized Singular Value Decomposition (SVD)~\cite{halko2011finding} and train separate linear regression models for each user on top of these embeddings. This separation enables parallelization across users and reduces the optimization to a convex objective that is globally optimized in closed-form~\cite{levy}. Unfortunately, naive SVD embedding methods often exhibit a popularity bias that skews their ability to accurately embed less popular items \cite{paterek2007improving}.


In this paper, we propose a novel projected linear recommendation algorithm called Noise Contrastive Estimation PLRec (NCE-PLRec).
Instead of explicitly treating unobserved interactions as
negative feedback, we leverage insights from the NCE framework~\cite{gutmann2010noise} that attempt to discriminate between observed interactions and a noise model; NCE has been previously used extensively in high-quality word embeddings for natural language~\cite{mikolov2013distributed,levy2014neural}.
Specifically, we first transform the implicit matrix into a de-popularized matrix that optimally re-weights the interactions in closed-form according to the NCE objective. 
Then we extract item embeddings by projecting items onto the principal components of this de-popularized matrix obtained via SVD.
We can then leverage the standard PLRec framework with these NCE item embeddings in a novel highly scalable OC-CF method that we call NCE-PLRec.

An analysis of recommendation popularity distribution demonstrates that NCE-PLRec uniformly distributes its recommendations over the popularity spectrum while other methods exhibit distinct biases towards specific popularity subranges. 
Overall, our results show that NCE-PLRec outperforms existing state-of-the-art models in terms of ranking metrics and results in very efficient training times for large-scale datasets such as Netflix and MovieLens-20m.



\section{Notation and Background} \label{sec:Background}
Before proceeding, we define some basic notation:
\begin{itemize}
\item $R$: $m \times n$ implicit feedback matrix. The entry of this matrix is either 1 (observed interaction) or 0 (no interaction). $\mathbf{r}_i$ represents all implicit feedback from user $i \in \{ 1\cdots m\}$, and $\mathbf{r}_{:,j}$ represents all user feedback for item $j \in \{1\cdots n\}$. We use $|\mathbf{r}_{:,j}|$ to represent the count of observed interactions for item $j$.
\item $U$, $V$: Latent representations (or embeddings) of users and items.  $U$ is $m \times k$, $V$ is $n \times k$.  We use $\mathbf{u}_i$ to represent the $i$th user representation (column of $U$), and $\mathbf{v}_j$ to represent the $j$th item representation (column of $V$).
\item $D = UV^T$: Inner product of user and item embeddings that has same shape of the implicit feedback matrix $R$.
\item $Q = R V$: $m \times k$ projected implicit matrix, obtained by projecting implicit matrix $R$ through item embeddings $V$.
\end{itemize}

\subsection{Matrix Factorization}
Matrix factorization models are a subset of latent factorization models, which attempt to uncover latent features of users and items that explain the observations in the implicit feedback matrix. It assumes the implicit feedback for a user and item is reconstructed through a function $g$ of the user's latent representation $u_i$ and item latent representation $v_j$:
\[\argmin_{U,V}\sum_{i,j}(r_{ij}-g(\mathbf{u}_{i}, \mathbf{v}_{j}))^2+\lambda(\norm{\mathbf{u}_i}_2^2+\norm{\mathbf{v}_j}_2^2+\norm{\theta}_2^2),\]
where function $g_\theta(\cdot)$ is often a dot product $\mathbf{u}_{i}^T\mathbf{v}_{j}$, or a complex neural network \cite{he2017neural}.

Matrix factorization (MF) methods~\cite{koren2009matrix} typically do not perform best in the OC-CF setting with implicit feedback data as they do not distinguish error contributions between positive and unobserved interactions.  Weighted Regularized MF (WRMF) \cite{hu2008collaborative} helps correct this by extending the original MF model by introducing a hyperparameter $\alpha$ to produce a term $c_{ij} = 1 + \alpha r_{ij}$ used to differentially scale the positive and negative error.  The first component of the objective above then takes the weighted form $\sum_{i,j}c_{ij}(r_{ij}-\mathbf{u}_{i}^T\mathbf{v}_{j})^2$ and $\alpha$ is tuned through cross-validation.


Although WRMF performs well in implicit feedback recommendation tasks, its training is inefficient and not parallelizable as it performs iterative updates using Alternating Least Squares (ALS) to minimize an upper bound of the reconstruction error. A slight variant uses Stochastic Gradient Descent (SGD) which is inefficient due to sampling negatives (i.e., sparsity cannot be exploited as it was in ALS).

\subsection{Linear Recommenders}
Compared to classic memory-based neighborhood models that compute a similarity function heuristically, linear models learn the similarity matrix directly via linear regression~\cite{sanner:aaai16b}.

Sparse LInear Method (SLIM) \cite{ning2011slim} learns the similarity matrix by minimizing the reconstruction error of the constrained objective function
\[\argmin_{S}\sum_{i,j}(r_{i,j}-\mathbf{r}_i\mathbf{w}_{j})^2+\frac{\beta}{2}\norm{\mathbf{w}_j}_2^2+\lambda\norm{\mathbf{w}_j}_1,\]
\[W\geq 0, \qquad diag(W)=0\]
where $W$ is the similarity matrix to be learned, and the constraints act as a regularizer to prevent the trivial solution where $W$ is the identity matrix $I$. 

Unfortunately, SLIM and its variants such as LRec~\cite{sanner:aaai16b} are not scalable as they require storing a large dense similarity matrix that grows quadratically with the number of users or items. This is impractical for real world problems with millions of users and items. Moreover, learning a large number of parameters relative to the sparse observations is ill-formed since it requires solving more unknowns than available equations.

Linear Flow~\cite{sanner:ijcai16a}, which we refer as Projected Linear Recommender (PLRec), addresses these issues by first reducing the dimensionality of the implicit matrix followed by Linear Regression:
\[\argmin_{W}\sum_{i,j}(r_{i,j}-\mathbf{r}_iV\mathbf{w}_{j}^T)^2+\lambda\norm{\mathbf{w}_j}_2^2,\]
where $V$ is item embedding matrix obtained from truncated SVD decomposition of the implicit matrix $R$; $R = U\Sigma V^T$. Since the item embedding dimension $k \ll min\{m, n\}$, training PLRec requires substantially fewer parameters to learn compared to SLIM. 

\subsection{Issues with Projected LRec}
PLRec faces two deficiencies that hurts its performance. Firstly, naively decomposing the implicit feedback matrix through an SVD decomposition directly makes the model highly biased to the large number of unobserved ratings. This results in underestimated item embeddings, especially for the unpopular items.  Secondly, since SVD is the optimal solution of decomposing the implicit matrix, the optimal $W$ learned from PLRec is exactly $V$ if regularization were ignored.  To understand this issue, we substitute $r_{i,j} = \mathbf{u}_i\Sigma\mathbf{v}_j$ in the PLRec objective function and obtain the following 
\[
\argmin_{W}\sum_{i,j}(r_{i,j}-\mathbf{u}_i\Sigma\mathbf{w}_{j})^2+\lambda\norm{\mathbf{w}_j}_2^2,
\]
since $V^TV = I$.  Empirically, PLRec often performs similar to the PureSVD~\cite{cremonesi2010performance} algorithm for this reason.

\subsection{Noise-Contrastive Estimation}
Noise-Contrastive Estimation (NCE)~\cite{gutmann2010noise} learns to discriminate between the observed data and some artificially generated noise. Given an observation dataset $X = \{\mathbf{x}_1\cdots \mathbf{x}_n\}$ and artificially generated dataset $Y = \{\mathbf{y}_1\cdots \mathbf{y}_n\}$, NCE maximizes an objective function that contrasts observations with noise:
\[\sum_j\log(g(\mathbf{x}_j; \theta)) + \log(1-g(\mathbf{y}_j;\theta)), \]
where $g(\cdot)$ is a (possibly unnormalized) logistic probability density function, and $\theta$ are model parameters to estimate.



\section{Noise Contrastive Item Embeddings}

So far, PLRec stands out as one of the most scalable OC-CF methods (as our results later verify), but it suffers from a popularity-biased embedding similar to methods such as PureSVD.  We now aim to leverage ideas from NCE to find an improved item embedding for PLRec.

We begin by revisiting recommendation from a probabilistic perspective, where we fit a model parameterized by the user and item embeddings to maximize the probability of observed feedback. Instead of explicitly treating unobserved interactions as negative feedback, NCE learns  properties of users and items in the form of a statistical model to discriminate between observed interactions and unobserved noise. 



\subsection{Noise Contrastive Estimation in Recommendation}
The probabilistic objective of recommendation is to train a model that maximizes the probability $p(r_{i,j}=1|i,j)$ for all positive observations of the user $i$ given the user embedding $\mathbf{u}_i$ and item embedding $\mathbf{v}_j$. Motivated by the log-odds ratio derived from a Bernoulli Distribution~\cite{banerjee2007analysis}, we define the rating probability as the sigmoid function: 
\begin{equation}
p(r_{i,j} = 1|i, j) = \sigma(\mathbf{u}_i^T\mathbf{v}_j) = \frac{1}{1 + e^{-\mathbf{u}_i^T\mathbf{v}_j}} .
\end{equation} 

Since the negative feedback is unobservable, we could artificially generate negative samples through sampling items from the item popularity $p(j') = \frac{|r_{:,j'}|}{\sum_{l'}^n |r_{:,l'}|}$ for each positive item rating of user $i$.
Thus, we are able to construct the following NCE objective for user $i$ where for each $j$ summand, we could sample an item $j'$ as negative noise according to its popularity $j' \sim p(j')$:
\begin{equation}
\begin{split}
&\argmax_{\mathbf{u}_i, V} \sum_j r_{i,j} \left [\log\sigma(\mathbf{u}_i^T\mathbf{v}_j) +\log\sigma(-\mathbf{u}_i^T\mathbf{v}_{j'})\right ].
\end{split}
\label{eq:objective}
\end{equation}

By the Law of Large Numbers, the infinite sampling of noisy negative converges to its expectation. Thus, the NCE objective contrasts the positive observation with the expectation of the noisy negative samples:
\begin{equation}
\begin{split}
&\argmax_{\mathbf{u}_i, V} \sum_j \!r_{i,j} \!\left [\log\sigma(\mathbf{u}_i^T\mathbf{v}_j)\! +\!E_{p(j')}[\log\sigma(-\mathbf{u}_i^T\mathbf{v}_{j'})]\right ].
\end{split}
\label{eq:objective}
\end{equation}

Intuitively, this preference objective assumes user $i$ prefers any observed item $j$ over all other unobserved items $j'$.

In the multi-user environment, the full objective function $\ell$ corresponds to a summation over each independent user, where the item embeddings are shared by all users:
\begin{equation}
\begin{split}
&\argmax_{U, V}\! \sum_i\! \sum_j\! r_{i,j}\! \left [\log\sigma(\mathbf{u}_i^T\mathbf{v}_j)\! +\! E_{p(j')}[\log\sigma(-\mathbf{u}_i^T\mathbf{v}_{j'})]\right ].
\end{split}
\label{eq:objective}
\end{equation}

Optimizing equation \ref{eq:objective} with respect to both user or item representations in closed-form is intractable due to shared item embeddings and nonlinear relation between user and item embeddings. Therefore, we optimize equation \ref{eq:objective} with respect to the dot product $d_{i,j} = \mathbf{u}_i^T\mathbf{v}_j$ directly to  simplify the objective into a convex optimization problem. Solving for the optimal $d_{i,j}$ for positive observations, we obtain
\begin{equation}
\begin{split}
\frac{\partial \ell }{\partial d_{i,j} } &= \sigma(-d_{i, j}) - \frac{|r_{:,j}|}{\sum_j' r_{:,j'}} \sigma(d_{i, j})
\end{split}
\end{equation}

\begin{equation}
d_{i,j}^* = \log \frac{\sum_{j'} |r_{:,j'}|}{|r_{:,j}|} \qquad \forall r_{i,j} = 1 .
\label{eq:5}
\end{equation}
For the unobserved interactions, the optimal solution is simply zero
\begin{equation}
d_{i,j}^*= 0 \qquad \forall r_{i,j} = 0.
\label{eq:6}
\end{equation}

The resulting sparse matrix $D$ maintains the same number of non-zero entries and shape from the original implicit matrix. The difference is that the entries are now replaced with the optimal inner product of user and item representations, $D^* = U^* V^{*T}$. 


Finally, we project the sparse $D^*$ using truncated SVD \cite{halko2011finding} as it exploits sparsity in the matrix.
\begin{equation}
\label{eq:optimalUV}
U^* \approx U_D\Sigma_D^{\frac{1}{2}}  \qquad V^* \approx V_D\Sigma_D^{\frac{1}{2}},
\end{equation}
where $U_D$, $V_D$ and $\Sigma_D$ are come from $D^* \approx U_D\Sigma_DV_D^T$.

\subsection{Relation to the Neural Word Embeddings}
Noise Contrastive Estimation was first brought to the attention of the Machine Learning community from the literature on word embeddings~\cite{mikolov2013efficient,mikolov2013distributed,levy2014neural,hashimoto2016word}.

Conceptually, our proposed objective is similar to word embeddings, where we analogize users as word contexts and items as words. The difference is that we assume the users (contexts) are unique and that the interactions (words) are independent with a uniform discrete distribution.



\subsection{NCE Item Embedding Hyperparameter}
The optimal solution of NCE as shown in Equation~\eqref{eq:5} penalizes the influence of popular items on the user and item representation. In other words, it is inversely proportional to the popularity of an observed item. However, this relies heavily on a good estimate of the popularity of observed items $p(j')$. Since the data is sparse, there is high uncertainty on the popularity estimate and this uncertainty propagates to Equation~\eqref{eq:5}.

To alleviate this, we introduce a hyperparameter $\beta$ into the denominator, which adjusts the penalty on high frequency items. We rewrite Equation~\eqref{eq:5} to include $\beta$ as follows: 
\begin{equation}
d_{i,j} = \max(\log \sum_{j'} |r_{:,j'}|-\beta\log|\mathbf{r}_{:,j}|,0) \quad \forall r_{i,j} = 1,
\label{eq:hyper_beta}
\end{equation}
where we add  a $d_{i,j} \geq 0$ constraint to guarantee the positive feedback is more significant compared to the unobserved feedback in Equation~\eqref{eq:6}. Empirically, this hyperparameter aids generalization on the test set as shown in Figure~\ref{fig:hyper-paramter}.

\subsection{Linear Collaborative Filtering}


Using optimal user $U^*$ and item $V^*$ embeddings from equation \ref{eq:optimalUV}, we can predict unobserved interactions with a simple dot product $U^*V^{*T}$. We refer to this method of performing NCE followed by truncated SVD as NCE-SVD. 

NCE-SVD effectively de-popularizes the dataset by rescaling entries in $R$ inversely proportional to their popularity. However, popularity bias can still be important in terms of ranking performance depending on the dataset~\cite{canamares2018should}. Therefore, following the approach of PLRec, we further perform Linear Regression on top of NCE-SVD for the model to learn the importance of different latent features for each user. We call this final solution, NCE-PLRec, as it performs NCE followed by PLRec \cite{sanner:aaai16b}.

In addition, the static latent representation of NCE-SVD is unable to capture drifting user preferences \cite{koren2009collaborative}. On the other hand, NCE-PLRec can adaptively train and update its weights according to the user's current taste.
To achieve this, we project the original implicit matrix $R$ onto the learned item representation $V^*$. This projection produces the dynamic user representation, $Q=RV^*$, which is the sum over all item representations of the user's interaction history.
Then, we maximize the conditional likelihood of observation $p(R|Q)$ as a Gaussian distribution:

\begin{equation}
\begin{aligned}
\argmin_{W}\sum_{i,j}c_{i,j}(r_{i,j}-\mathbf{q}_i\mathbf{w}_{j})^2+\lambda\norm{\mathbf{w}_j}_2^2,
\end{aligned}
\label{eq:hyper_alpha}
\end{equation}
where $\mathbf{q}_i$ is the dynamic user representation, $c_{i,j}$ is the loss weighting hyper-parameter, and $W$ is the linear regression coefficient matrix to estimate.

The definition of the weighting matrix $C$ above is:
\begin{equation}
c_{i,j} = 1+ \alpha r_{i,j}
\label{eq: weightMatrix}
\end{equation}
where we constrain the hyper-parameter value to be $\alpha \geq -1$

\subsection{Prediction on Cold-Start Test Users}
Referring to Equation~\eqref{eq:hyper_alpha}, the trained weights $\mathbf{w}_{j}$ for each item $j$ are shared and trained by all train users. 
Given a cold-start test user, whose ratings were not used during training, $\mathbf{r}_{i'}$, we recommend the top-K items from its projection onto the item features and weights learned from the train users $\mathbf{r}_{i'} V^*W^T = \mathbf{q}_{i'}W^T$.

\subsection{Algorithm}
We summarize the Noise-Contrastive Estimation Projected Linear Recommender (NCE-PLRec) in Algorithm \ref{alg:wrelr}.
\begin{algorithm}
\caption{NCE-PLRec}\label{alg:wrelr}
\begin{algorithmic}[1]
\Procedure{Train}{$R$, $\alpha=0$, $\beta=1$}
\State $D^*\gets \textit{NCE}(R, \beta)$ \Comment{Construct D matrix}
\State $U_D$,$\Sigma_D$,$V_D^T \gets \textit{Truncated SVD}(D^*)$ 
\State $Q \gets RV_D\Sigma_D^{\frac{1}{2}}$ \Comment{Project  implicit matrix}
\For{$i \in range(1,m)$}\Comment{Loop over users}
\State $C^j \gets \textit{diag}(1 + \alpha \mathbf{r}_{:,j})$
\State $\mathbf{w}_j \gets ( Q^TC^jQ +\lambda I)^{-1}Q^TC^j\mathbf{r}_{:,j}$
\EndFor\label{euclidendwhile}
\State \textbf{return} $QW^T$ \Comment{Prediction}
\EndProcedure
\end{algorithmic}
\end{algorithm}

The optimization appears to be closed-form only with respect to each user. However, if the weighting hyper-parameters is fixed at $\alpha=0$ for the observed entries as done in Equation~\eqref{eq: weightMatrix}, it simplifies into a globally closed-form solution, $W = (Q^TQ +\lambda I)^{-1}Q^TR$ . The globally closed-form (NCE-PLRec) executes faster than the per user closed-form (NCE-PLRec-W) as shown in Figure \ref{fig:time}.

\begin{table*}[t!]
\caption{Results of Movielens-20M dataset with 95\% confidence interval. Hyper-parameters are chosen from the validation set.}
\rowcolors{2}{gray!10}{white}
\resizebox{\textwidth}{!}{%
\begin{tabular}{lllllllllll}
\toprule
model&NDCG&Precision@5&Precision@10&Precision@20&R-Precision&Recall@5&Recall@10&Recall@20\\
\midrule
POP&0.1194$\pm$0.0007&0.0945$\pm$0.0009&0.0854$\pm$0.0007&0.0751$\pm$0.0006&0.068$\pm$0.0005&0.0327$\pm$0.0004&0.0581$\pm$0.0006&0.0974$\pm$0.0008\\
PLRec&0.1622$\pm$0.0007&0.1213$\pm$0.0009&0.1105$\pm$0.0007&0.0976$\pm$0.0006&0.089$\pm$0.0005&0.042$\pm$0.0005&0.0754$\pm$0.0006&0.1277$\pm$0.0009\\
PureSVD&0.1615$\pm$0.0007&0.1207$\pm$0.0009&0.11$\pm$0.0007&0.0972$\pm$0.0006&0.0884$\pm$0.0005&0.0416$\pm$0.0005&0.0749$\pm$0.0006&0.1273$\pm$0.0009\\
WRMF&0.1832$\pm$0.0008&0.1213$\pm$0.0009&0.1127$\pm$0.0007&0.1028$\pm$0.0006&0.0929$\pm$0.0006&0.0451$\pm$0.0005&0.0823$\pm$0.0007&0.1476$\pm$0.001\\
AutoRec&0.1697$\pm$ 0.0009&0.1288$\pm$ 0.001&0.1163$\pm$ 0.0008&0.1019$\pm$ 0.0006&0.0929$\pm$ 0.0006&0.0452$\pm$ 0.0005&0.0791$\pm$ 0.0007&0.1308$\pm$ 0.001\\
CML&0.1755$\pm$0.0008&0.1191$\pm$0.001&0.1125$\pm$0.0008&0.1036$\pm$0.0006&0.0897$\pm$0.0006&0.0392$\pm$0.0005&0.0741$\pm$0.0007&0.1363$\pm$0.0009\\
\midrule
NCE-SVD&0.1553$\pm$0.0007&0.1067$\pm$0.0009&0.0984$\pm$0.0007&0.089$\pm$0.0005&0.079$\pm$0.0005&0.0392$\pm$0.0005&0.0726$\pm$0.0007&0.1311$\pm$0.0009\\
NCE-PLRec&{\bf 0.1968$\pm$0.0009}&{\bf 0.1407$\pm$0.0011}&{\bf 0.1282$\pm$0.0008}&{\bf 0.1143$\pm$0.0007}&{\bf 0.103$\pm$0.0006}&{\bf 0.0497$\pm$0.0005}&{\bf 0.0894$\pm$0.0007}&{\bf 0.1565$\pm$0.001}\\

\bottomrule
\end{tabular}}
\label{table:ml}
\end{table*}

\begin{table*}[t!]
\caption{Results of Netflix dataset with 95\% confidence interval. Hyper-parameters are chosen from the validation set.}
\rowcolors{2}{gray!10}{white}
\resizebox{\textwidth}{!}{%
\begin{tabular}{lllllllllll}
\toprule
model&NDCG&Precision@5&Precision@10&Precision@20&R-Precision&Recall@5&Recall@10&Recall@20\\
\midrule
POP&0.0853$\pm$0.0003&0.0711$\pm$0.0004&0.0709$\pm$0.0003&0.0663$\pm$0.0003&0.0486$\pm$0.0002&0.0179$\pm$0.0002&0.0301$\pm$0.0002&0.0532$\pm$0.0003\\
PLRec&0.1554$\pm$0.0004&0.1474$\pm$0.0005&0.1317$\pm$0.0004&0.115$\pm$0.0003&0.0948$\pm$0.0003&0.0421$\pm$0.0003&0.0703$\pm$0.0003&0.1135$\pm$0.0004\\
PureSVD&0.1545$\pm$0.0004&0.1473$\pm$0.0005&0.1314$\pm$0.0004&0.1146$\pm$0.0003&0.0944$\pm$0.0003&0.0417$\pm$0.0003&0.0698$\pm$0.0003&0.1126$\pm$0.0004\\
WRMF&0.1637$\pm$0.0004&0.1365$\pm$0.0005&0.1265$\pm$0.0004&0.1139$\pm$0.0003&0.0979$\pm$0.0003&0.045$\pm$0.0003&0.0773$\pm$0.0004&0.1268$\pm$0.0005\\
AutoRec&0.1491$\pm$0.0004&0.1225$\pm$0.0004&0.1334$\pm$0.0005&0.109$\pm$0.0003&0.0894$\pm$0.0003&0.0376$\pm$0.0003&0.0653$\pm$0.0003&0.1079$\pm$0.0004\\
CML&0.1487$\pm$0.0004&0.1307$\pm$0.0005&0.1212$\pm$0.0004&0.1091$\pm$0.0003&0.0865$\pm$0.0003&0.0354$\pm$0.0002&0.0638$\pm$0.0003&0.1104$\pm$0.0004\\
\midrule
NCE-SVD&0.1553$\pm$0.0004&0.1491$\pm$0.0006&0.1317$\pm$0.0004&0.113$\pm$0.0003&0.0911$\pm$0.0003&0.0437$\pm$0.0003&0.0743$\pm$0.0004&0.1207$\pm$0.0005\\
NCE-PLRec&{\bf 0.1776$\pm$0.0004}&{\bf 0.1609$\pm$0.0006}&{\bf 0.1459$\pm$0.0005}&{\bf 0.1276$\pm$0.0004}&{\bf 0.1074$\pm$0.0003}&{\bf 0.0496$\pm$0.0003}&{\bf 0.0842$\pm$0.0004}&{\bf 0.1359$\pm$0.0005}\\
\bottomrule
\end{tabular}}
\label{table:netflix}
\end{table*}

\section{Experiments and Evaluation}
In this section, we evaluate the proposed NCE-PLRec model by comparing to a list of state-of-the-art OC-CF algorithms on three real-world datasets with at least 10 million interactions. The comparison includes general Top-K recommendation performance, time consuming, and popularity item sensitivity. 

We ran our experiments on a single Ubuntu Linux system workstation with one AMD Ryzen3 1400 4 core CPU, 16GB RAM, and one GTX 1070 GPU. Implementation is done with Python 2.7 and includes Tensorflow 1.4 \cite{abadi2016tensorflow}.  Code to reproduce results is available on Github.\footnote{\url{https://github.com/wuga214/NCE_Projected_LRec}}

\subsection{Datasets}
We evaluate the candidate algorithms on three publicly available
rating datasets: Movielens-20M, Netflix Prize, and Yahoo R1. Each dataset contains more than 10 million interactions. Thus, we are only able to compare with state-of-the-art models that are able to run on these large-scale datasets. For each dataset, we binarize the rating dataset with a rating threshold, $\vartheta$, defined to be the upper half of the range of the ratings. We do this so that the observed interactions correspond to positive feedback. To be specific,  the threshold is $\vartheta > 3$ for Movielens-20M and Netflix Prize, and $\vartheta > 50$ for Yahoo R1. Table \ref{tb:dataset} summarizes the properties of the binarized matrices. 

\begin{table}[h!]
\caption{Summary of datasets used in evaluation.}
  \rowcolors{2}{gray!10}{white}
  \resizebox{\linewidth}{!}{%
  \begin{tabular}{ccccc}
  \toprule
    Dataset & $m$ & $n$ & $|r_{i,j}>\vartheta|$& Sparsity\\
   \midrule
    MovieLens-20m & 138,493&27,278&12,195,566&$3.47\times10^{-3}$\\
    Netflix Prize & 2,649,430& 17,771 & 56,919,190&$1.2\times10^{-3}$ \\
    YahooR1 Data & 1,948,882 & 46110 & 61,335,886 &$6.82\times10^{-4}$\\
    \hline
  \end{tabular}
  }
 \label{tb:dataset}
\end{table}

We split the data into train, validation and test sets based on timestamps given by the dataset if they are available as it is more realistic \cite{cremonesi2010performance}. For each user, we use the first 50\% of data as the train set, 20\% data as validation set and 30\% data as the test set. For the Yahoo dataset, we split the dataset randomly as it does not contain timestamps. 

\subsection{Evaluation Metrics}
We evaluate the recommendation performance using four metrics: Precision@K, Recall@K, R-Precision, and B-NDCG, where R-Precision is an order insensitive metrics, NDCG is order sensitive, and Precision@K as well as Recall@K are semi-order sensitive due to the K values given.






\subsection{Candidate Methods}
We compare the proposed algorithm with six seminal models from classical matrix factorization to the latest Collaborative Metric Learning. These models are scalable and run within reasonable time.
\begin{itemize}
\item POP: Most popular items -- not user personalized but an intuitive baseline to test the claims of this paper.
\item PLRec \cite{sanner:ijcai16a}: Also called Linear-Flow. This is the baseline projected linear recommendation approach. We run 7 truncated SVD iterations to guarantee the model converges.  This is one ablation of NCE-PLRec.
\item PureSVD \cite{cremonesi2010performance}: A similarity based recommendation method that constructs a similarity matrix through SVD decomposition of implicit matrix $R$.  
\item WRMF \cite{hu2008collaborative}: Weighted Regularized Matrix Factorization as described previously. We run 7 alternating least squares iterations to convergence.
\item AutoRec \cite{sedhain2015autorec}: A neural Autoencoder based recommendation system with one hidden layer and Relu activation function. We train for $200$ epochs until training convergence is achieved.
\item CML \cite{hsieh2017collaborative}: Collaborative Metric Learning. A state-of-the-art metric learning based recommender system. $20$ iterations reaches training convergence.
\item NCE-SVD: Inner product of SVD-decomposed item and user representation learned from NCE.  This is an ablation of NCE-PLRec without PLRec's learned linear models.
\item NCE-PLRec: The full version of the proposed model.
\end{itemize}

\begin{table}[h!]
 \caption{Hyperparameters tuned on the experiments.}
  \rowcolors{2}{gray!10}{white}
  \resizebox{\linewidth}{!}{%
  \begin{tabular}{cccc}
  \toprule
    name & Range & Functionality& Algorithms affected \\
   \midrule
    k & \{50, 100, 200, 500\}& Latent Dimension&\begin{tabular}{@{}l@{}}PLRec, PureSVD \\ WRMF, AutoRec, CML\\NCE-SVD, NCE-PLRec \end{tabular}\\
    $\alpha$ & \begin{tabular}{@{}l@{}}\{-0.5, -0.4 $\cdots$ -0.1\} $\cup$\\ \{0, 0.1, 1, 10, 100\}\end{tabular}& Loss Weighting& WRMF, NCE-PLRec\\
    $\beta$ & \{0.7, 0.8 $\cdots$ 1.3\}& Popularity Sensitivity& NCE-SVD, NCE-PLRec\\
    $\lambda$ & \{0.001, 0.01, 0.1, 1, 10, 100\}& Regularization&\begin{tabular}{@{}l@{}}PLRec, WRMF \\ AutoRec, CML\\NCE-PLRec \end{tabular}\\
    \hline
  \end{tabular}
  }%
 \label{tb:hyper-parameter}
\end{table}

\begin{table*}[t!]
\caption{Results of Yahoo dataset with 95\% confidence interval. Hyperparameters are chosen from the validation set.}
\rowcolors{2}{gray!10}{white}
\resizebox{\textwidth}{!}{%
\begin{tabular}{lllllllllll}
\toprule
model&NDCG&Precision@5&Precision@10&Precision@20&R-Precision&Recall@5&Recall@10&Recall@20\\
\midrule
POP&0.1635$\pm$0.0002&0.0632$\pm$0.0002&0.0567$\pm$0.0001&0.0481$\pm$0.0001&0.0612$\pm$0.0002&0.0625$\pm$0.0002&0.1095$\pm$0.0003&0.1794$\pm$0.0004\\
PLRec&0.1967$\pm$0.0003&0.1671$\pm$0.0003&0.1397$\pm$0.0002&0.1105$\pm$0.0002&0.1289$\pm$0.0002&0.1042$\pm$0.0002&0.1631$\pm$0.0003&0.2412$\pm$0.0003\\
PureSVD&0.1943$\pm$0.0003&0.1652$\pm$0.0003&0.1379$\pm$0.0002&0.1094$\pm$0.0002&0.1275$\pm$0.0002&0.1034$\pm$0.0002&0.1612$\pm$0.0003&0.2382$\pm$0.0003\\
WRMF&0.2572$\pm$0.0003&0.2028$\pm$0.0003&0.1738$\pm$0.0002&0.1395$\pm$0.0002&0.165$\pm$0.0002&0.137$\pm$0.0003&0.2214$\pm$0.0003&0.3285$\pm$0.0004\\
AutoRec&0.1697$\pm$0.0002&0.0707$\pm$0.0002&0.0608$\pm$0.0001&0.0499$\pm$0.0001&0.066$\pm$0.0002&0.0718$\pm$0.0002&0.1185$\pm$0.0003&0.184$\pm$0.0004\\
CML&0.2841$\pm$0.0003&0.1214$\pm$0.0002&0.1095$\pm$0.0002&0.0937$\pm$0.0001&0.1146$\pm$0.0002&0.1147$\pm$0.0003&0.1998$\pm$0.0004&0.3266$\pm$0.0004\\

\midrule
NCE-SVD&0.1811$\pm$0.0003&0.0949$\pm$0.0002&0.0782$\pm$0.0001&0.0619$\pm$0.0001&0.0903$\pm$0.0002&0.0880$\pm$0.0003&0.1344$\pm$0.0003&0.1964$\pm$0.0004\\
NCE-PLRec&{\bf 0.3348$\pm$0.0003}&{\bf 0.2585$\pm$0.0003}&{\bf 0.2242$\pm$0.0003}&{\bf 0.1805$\pm$0.0002}&{\bf 0.2146$\pm$0.0003}&{\bf 0.1781$\pm$0.0003}&{\bf 0.2908$\pm$0.0004}&{\bf 0.4331$\pm$0.0004}\\
\bottomrule
\end{tabular}}
\label{table:yahoo}
\end{table*}

\begin{figure*}[h!]
\centering
	\begin{subfigure}{0.243\textwidth}
  	\includegraphics[width=1\linewidth]{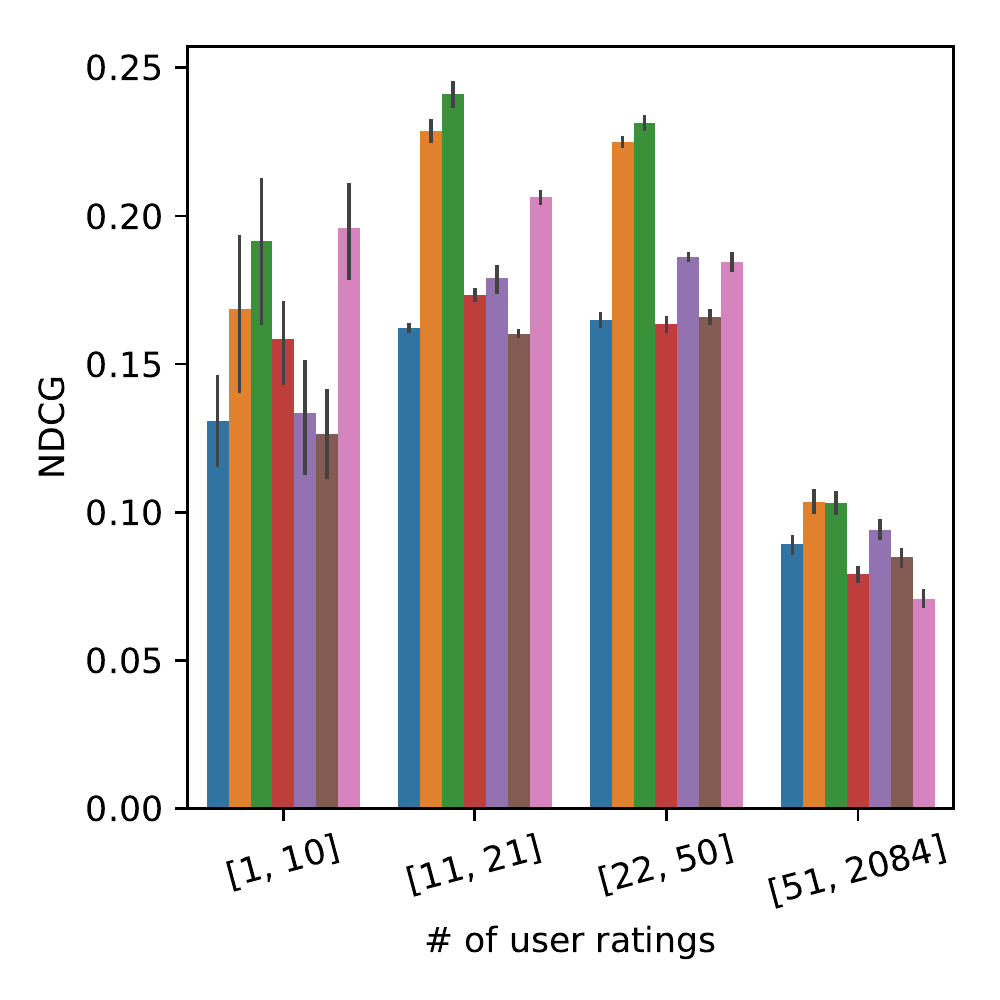}
    \caption{NDCG}
  	\end{subfigure}
	\begin{subfigure}{0.243\textwidth}
  	\includegraphics[width=1\textwidth]{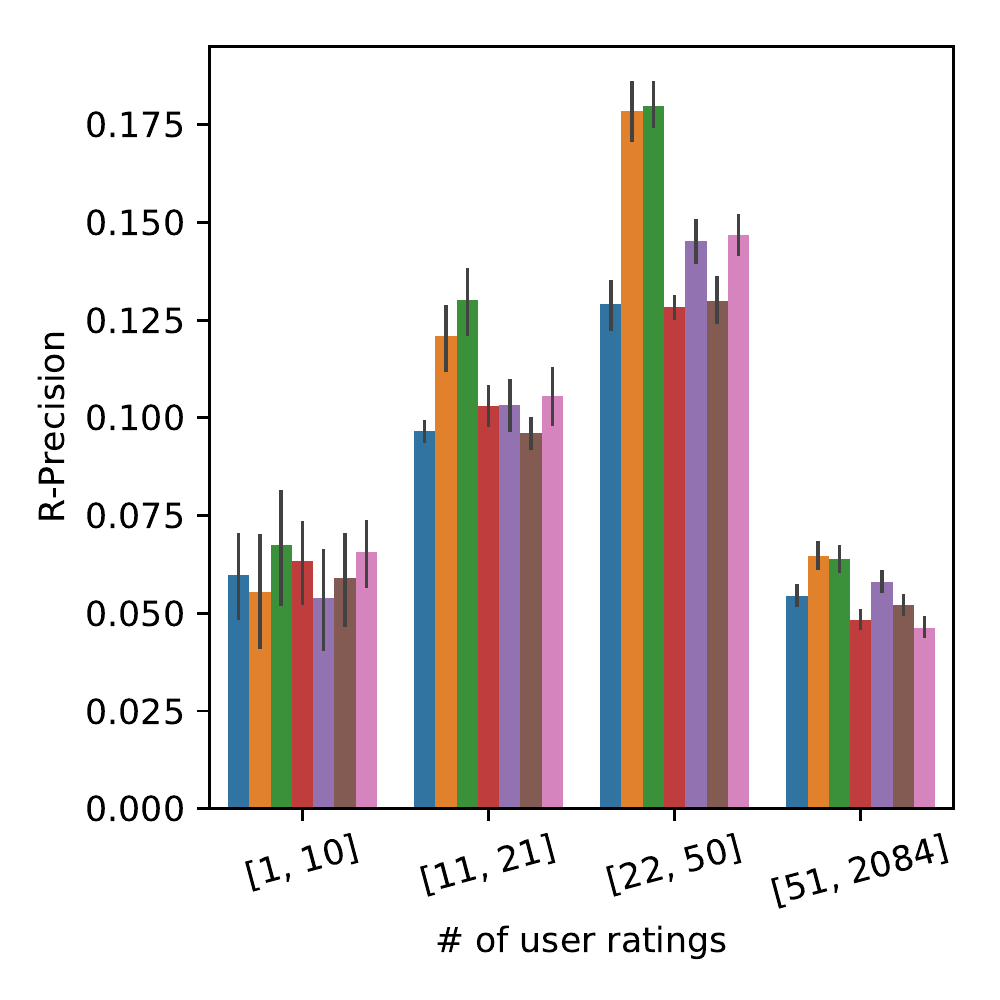}
    \caption{R-Precision}
    \end{subfigure}
    \begin{subfigure}{0.243\textwidth}
  	\includegraphics[width=1\linewidth]{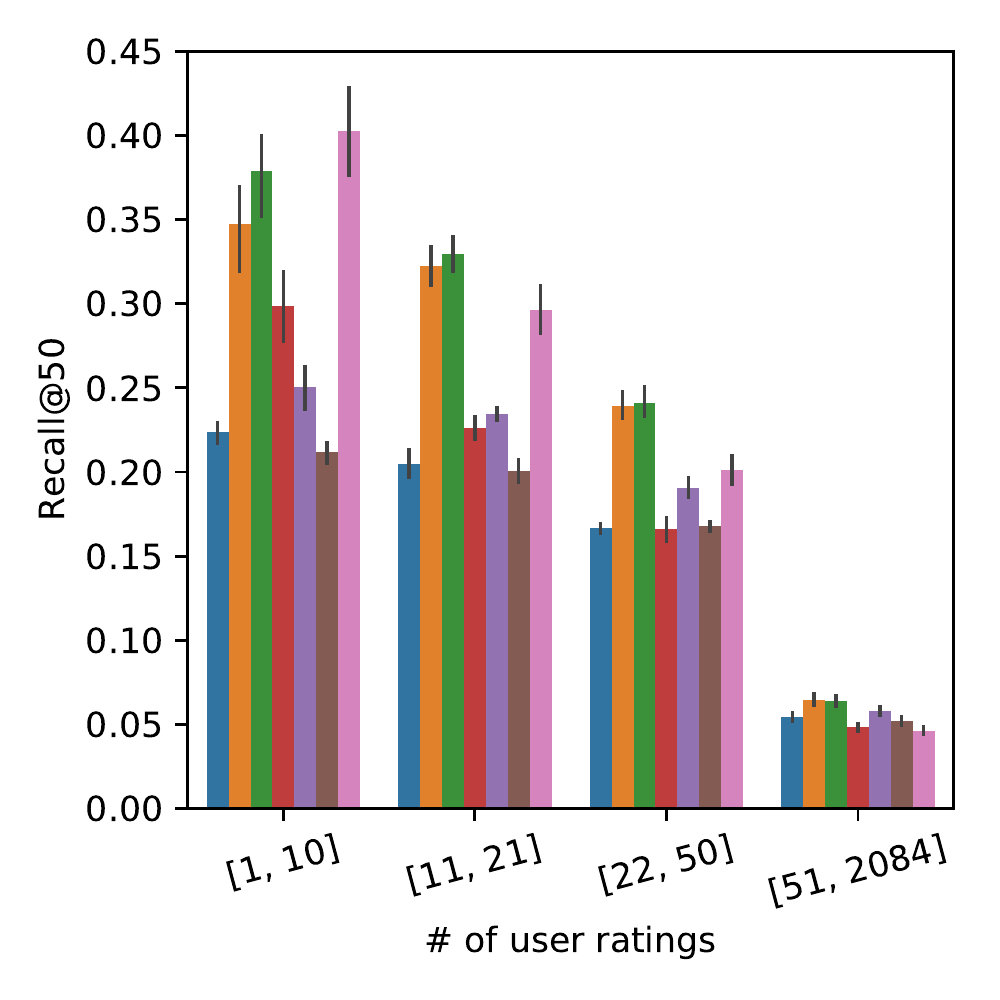}
    \caption{Recall@50}
  	\end{subfigure}
    \begin{subfigure}{0.243\textwidth}
  	\includegraphics[width=1\linewidth]{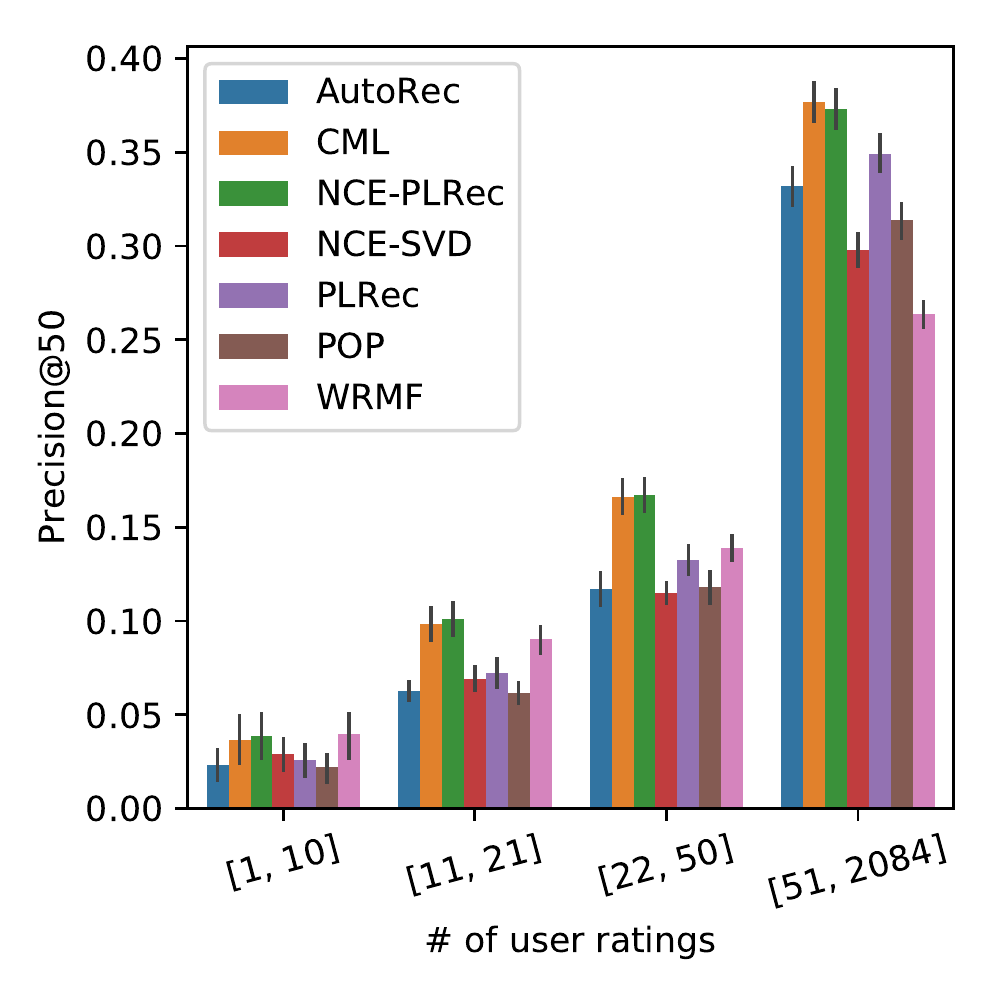}
    \caption{Precision@50}
  	\end{subfigure}
  \caption{Performance comparison for different user categories. Error bar show standard derivation. All figures share the legend.}
  \label{fig:user_catagorical_comparison}
\end{figure*}



\subsection{Ranking Performance Evaluation}
Tables \ref{table:ml}, \ref{table:netflix} and \ref{table:yahoo} show the general performance comparison between the proposed model with the six existing methods on all metrics. The best hyperparameters are learned through grid search as shown in Table \ref{tb:hyper-parameter}. From the results, we notice the following observations: (a) The proposed NCE-PLRec model outperforms all six candidate methods on all metrics in the experiments. It shows a substantial performance improvement compared to PLRec. (b) Predicting using NCE-SVD directly is not competitive because it is de-popularized. (c)~WRMF is the strongest competitor in terms of general performance, which reflects its wide use in practice. (d) CML is inconsistent as it performs well on Movielens-20m and Yahoo R1, but performs poorly on Netflix. (e)~PLRec and PureSVD show similar performances across all three datasets. This observation supports our theoretical claim that PLRec should learn a near-optimal weight $W \approx V$ from the SVD decomposition. 

\subsection{Performance vs. User Interaction Level} 

We now investigate conditions where the proposed algorithm works better compared to the strongest baselines. 
We categorize users based on the number of interactions they made in the training set into 4 categories. The categories come from the 25\%, 50\%, 75\%, and 100\% quantiles of the number of training interactions, which indicate how often the user rated items in the training set.

Figure~\ref{fig:user_catagorical_comparison} shows the comparison results in regard to the four categories. In general, NCE-PLRec shows strong performance for all the four rating distributions of users. CML shows competitive performance when the number of observed ratings are larger than 11. This is reasonable because CML requires more observations to effectively estimate the relative distance. WRMF also shows robust performance over the user categories with a lower number of ratings. Specifically, it outperforms NCE-PLRec when the number of ratings are less than 10. The reason for the poor performance of WRMF with a dense number of ratings has been studied thoroughly~\cite{hsieh2017collaborative}.

We also observed a strong alignment between NCE-PLRec and CML, which suggests that NCE-PLRec approximates \textit{metric learning} approaches. We investigate the reason for such alignment and observe that the objective function in Equation~\eqref{eq:objective} of NCE can be equated to unnormalized Cosine Distance Metric Learning, which maximizes the unnormalized cosine similarity between users and their observed items while minimizing the unobserved items.

\begin{figure}[h!]
\centering
\hspace{-2mm}\includegraphics[width=1.0\linewidth]{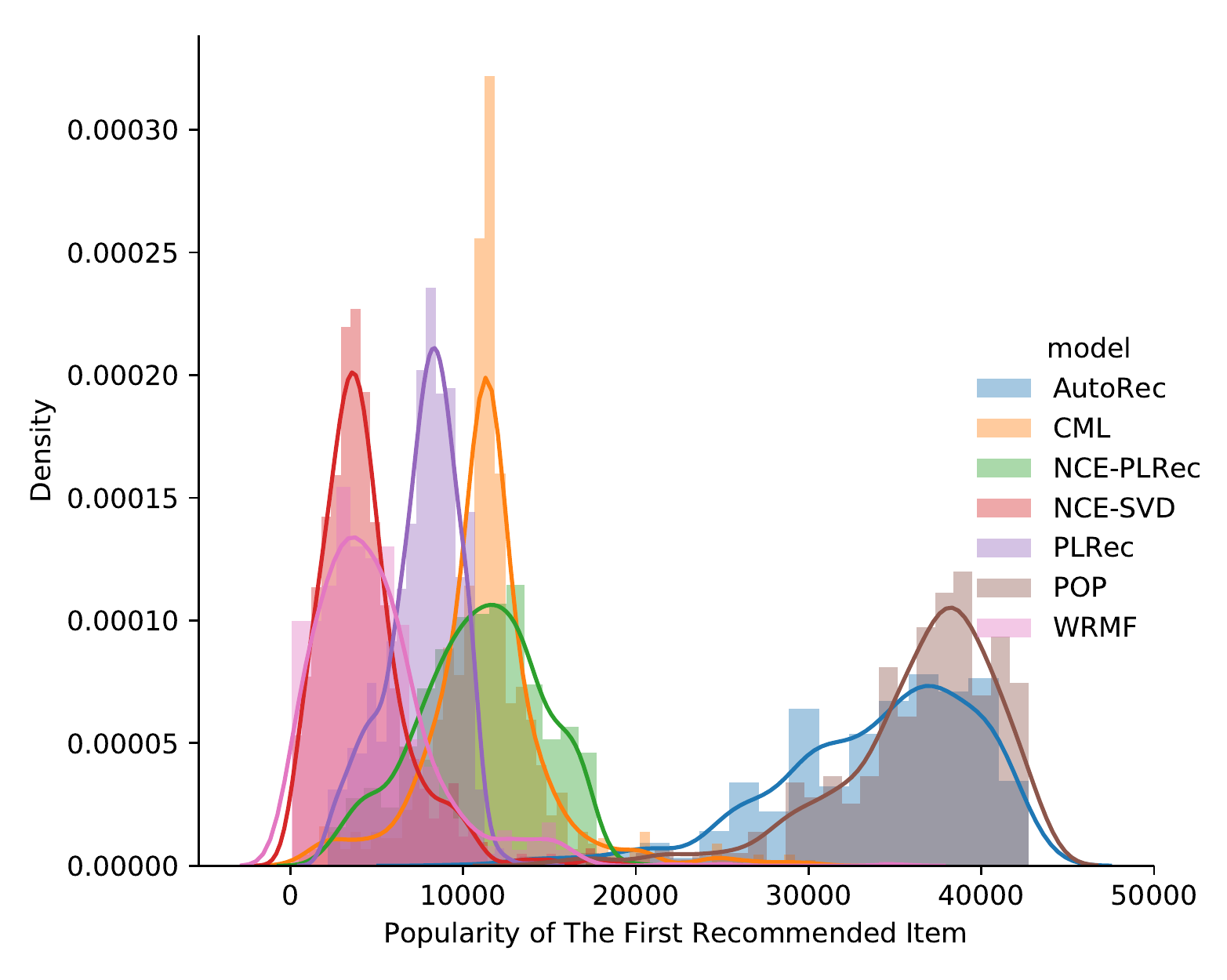}
\caption{Popularity of the first item recommended by all candidate algorithms. Note: PureSVD fully overlaps with PLRec and is not shown to reduce graph clutter.}
\label{fig:popularity}
\end{figure}

\subsection{Popularity Distribution of Recommendations}
We analyze the sensitivity of the candidate methods recommendations on popular items as shown in Figure \ref{fig:popularity}. In general, most of the candidate learning methods show strong personalization of recommendations except AutoRec, which tends to recommend popular items. On the other hand, NCE-SVD learns to only recommend unpopular items since the NCE embedding is de-popularized.  Impressively, NCE-PLRec
spreads its recommendations over the popularity spectrum compared to other algorithms and this proves to be beneficial in terms of its overall ranking performance previously observed in Tables~\ref{table:ml}, \ref{table:netflix} and~\ref{table:yahoo}.

\begin{figure}[h!]
\centering
\hspace{-5mm}\includegraphics[width=0.9\linewidth]{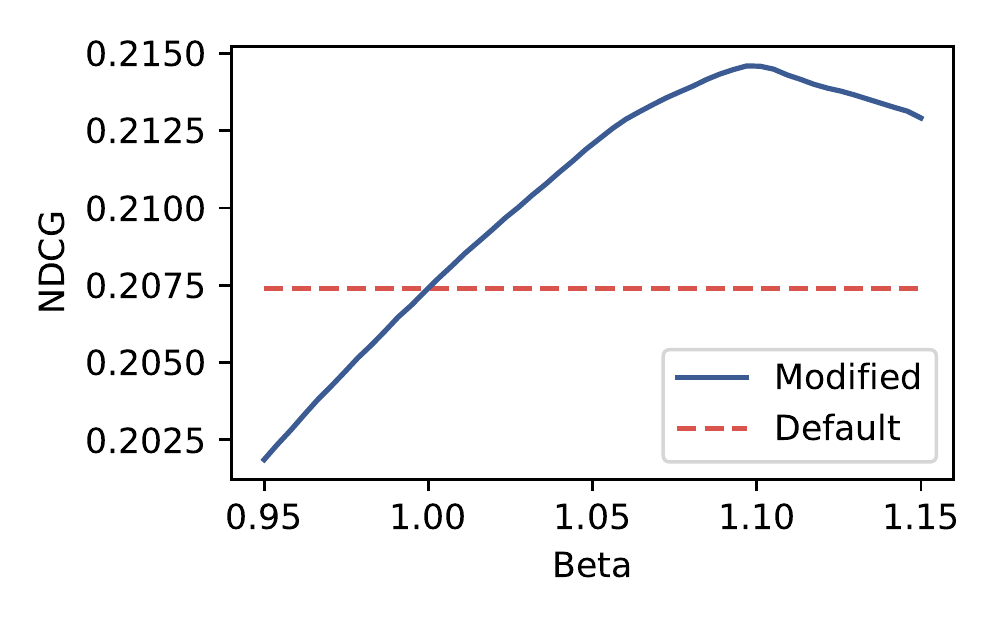}
\caption{Tuning the hyperparameter $\beta$ on Movielens-20m dataset. Blue curve shows the performance of tuning, while red dashed line shows the performance with default $\beta=1$.}
  \label{fig:hyper-paramter}
\end{figure}

\subsection{Hyperparameter Tuning}
Figure \ref{fig:hyper-paramter} shows the effects of tuning hyper-parameter $\beta$ for NCE-PLRec defined in Equation~\eqref{eq:hyper_beta} on NDCG in the Movielens-20m dataset (performance on other metrics was similar).  We observe a remarkable performance improvement by adjusting the weighting of the noise contrastive term.  This observation corresponds to our conjecture that this adjustment of the level of depopularization may be critical for working with extremely sparse recommendation data.

\begin{figure}[h!]
\centering
\hspace{-3mm}\includegraphics[width=0.9\linewidth]{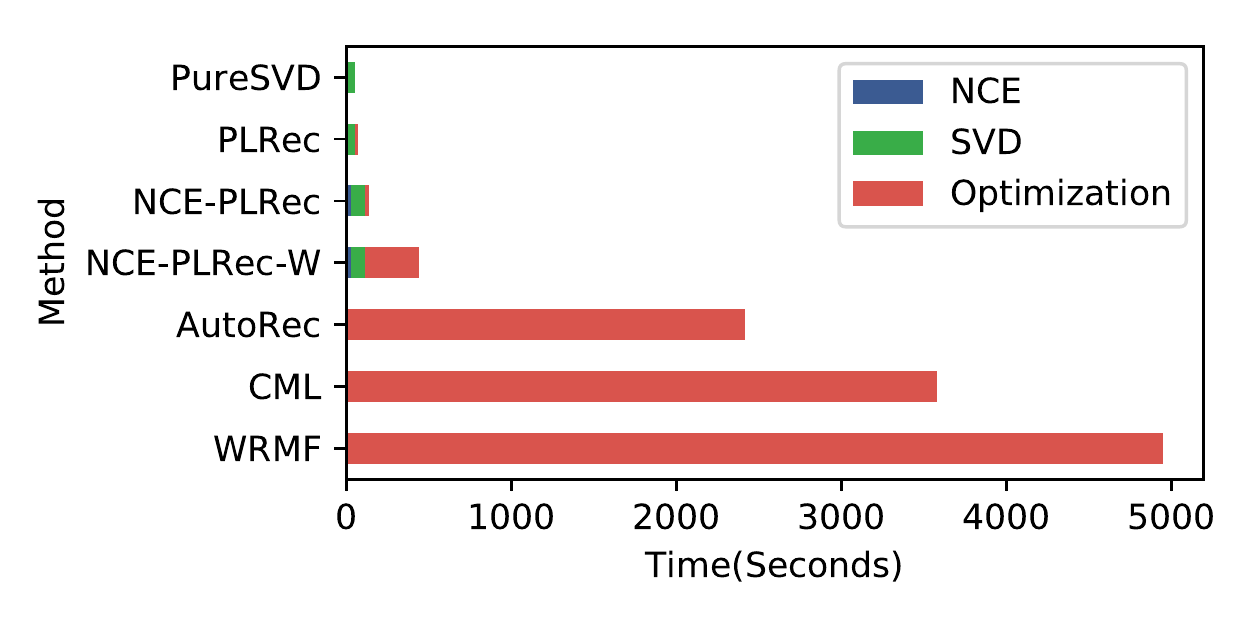}
\caption{Training times in seconds of the various methods on Movielens-20m. NCE-PLRec-W represents the model with loss weighting.}
  \label{fig:time}
\end{figure}

\subsection{Training Time and Scalability}

Figure~\ref{fig:time} shows the total time taken for training the candidate methods on the Movielens-20m dataset. We compare only the training time since the prediction and evaluation step require similar operations for all algorithms and take approximately the same time. The result shows the significant efficiency improvements from the linear models compared to neural network and alternating least squares training.  All PLRec methods including NCE-PLRec easily scale to these very large datasets.


\subsection{Cold-Start Test Users Case Study}
Among the recommendation methods, PureSVD, PLRec and NCE-PLRec are able to handle cold-start recommendations without leveraging additional side information. Since PLRec and PureSVD behave similarly, we only compare NCE-PLRec to PLRec for our user cold-start case study.

Due to limited space, Table \ref{tb:cherrypick} shows just two examples of cold-start recommendations, where we randomly create two test users that were not used during training.  While PLRec and NCE-PLRec overlap somewhat in their recommendations, we note that where they differ, NCE-PLRec appears to have chosen slightly more niche movies.

\begin{table}[t!]
\caption{Example of Cold-start Recommendation}
  \rowcolors{2}{gray!10}{white}
  \resizebox{\linewidth}{!}{%
  \begin{tabular}{lll}
  \toprule
    User Ratings & PLRec Recommendation & NCE-PLRec Recommendation\\
   \midrule
    \begin{tabular}{@{}l@{}}A Time to Kill (1996) \\ My Own Private Idaho (1991) \end{tabular} & 
    \begin{tabular}{@{}l@{}}Titanic (1997) \\ Fried Green Tomatoes (1991)\\A Few Good Men (1992)\end{tabular} &
    \begin{tabular}{@{}l@{}}A Few Good Men (1992) \\ Good Will Hunting (1997)\\  Philadelphia (1993) \end{tabular}\\
    \midrule
    \begin{tabular}{@{}l@{}}Three Colors: White (1994) \\ Six of a Kind (1934)\\Mina Tannenbaum (1994)\\The Love Letter (1999) \end{tabular} & 
    \begin{tabular}{@{}l@{}}Three Colors: Red (1994) \\ The Shawshank Redemption (1994)\\ Secrets \& Lies (1996)\end{tabular} &
    \begin{tabular}{@{}l@{}}Three Colors: Red (1994) \\ Secrets \& Lies (1996)\\Like Water for Chocolate (1992)\end{tabular}\\
    \hline
  \end{tabular}
  }
 \label{tb:cherrypick}
\end{table}

\begin{figure}[t!]
\centering
\hspace{-3mm}\includegraphics[width=0.6\linewidth]{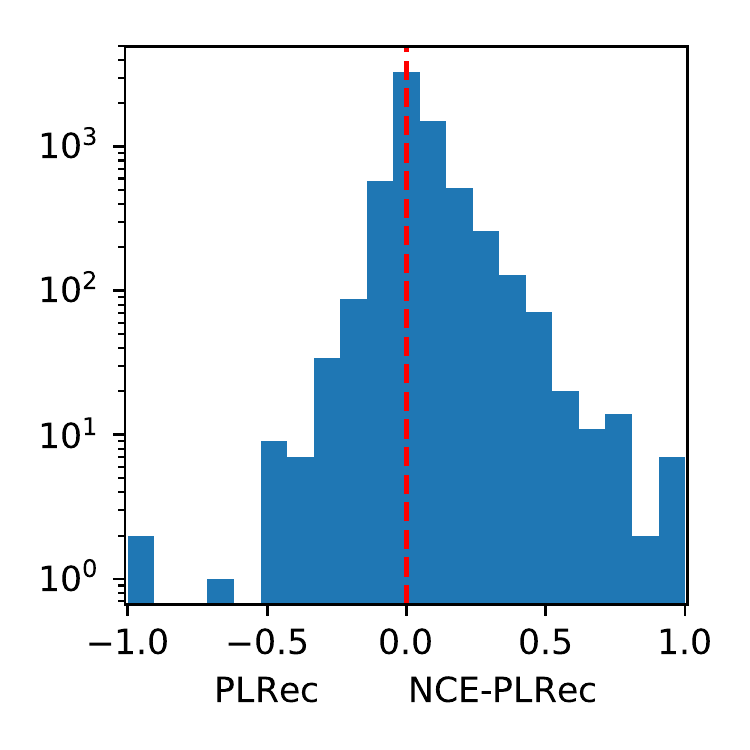}
\caption{Cold-start comparison histogram for Recall@50 of NCE-PLRec minus Recall@50 for PLRec. Positive values show NCE-PLRec has higher Recall whereas negative shows PLRec has higher Recall. The significant skew of area to the right side of the dotted 0.0 red line indicates that more cold-start users benefited from NCE-PLRec.}
  \label{fig:cold_hist}
\end{figure}


Figure \ref{fig:cold_hist} shows a more comprehensive pairwise comparison between NCE-PLRec and PLRec for the cold-start test users evaluation. In this experiment, we randomly remove 5\% of the users from the training dataset and use the remaining users for training. Then, we use the trained model to recommend items to the 5\% of held-out cold-start test users and evaluate performance. Clearly, most of the users received better cold-start recommendations from NCE-PLRec compared to PLRec in terms of Recall@50.

\section{Conclusion}
We proposed a novel linear recommendation algorithm called Noise Contrastive Estimation Projected Linear Recommendation (NCE-PLRec) that leverages item embeddings learned from NCE to make predictions using the highly scalable PLRec approach. We showed that NCE-PLRec outperforms several robust and scalable recommendation methods in almost all metrics.  Furthermore, NCE-PLRec is highly efficient during training, personalized with little popularity bias, and able to effectively handle cold-start user recommendation without leveraging side information.

\bibliographystyle{aaai}
\bibliography{aaai18}


\end{document}